# INVESTIGATING THE HYDRATION OF CM2 METEORITES BY IR SPECTROSCOPY.

S. Góbi[1], Á. Kereszturi[1], P. Beck[2], E. Quirico[2], B. Schmidt[2]
[1] Konkoly Observatory, Hungarian Academy of Sciences, H-1121 Budapest, Hungary,
e-mail: gobi.sandor@csfk.mta.hu
[2] Institut de Planétologie et d'Astrophysique de Grenoble, F-38400 Saint Martin d'Héres, France

**Introduction.** Chondritic meteorites are of great interest since they are one of the most ancient remnants of the early solar system. Some of them, like the carbonaceous CM meteorites experienced aqueous alteration thus their olivine content transformed more or less into hydrated silicates such as phyllosilicates. These hydrated CM2 meteorites have been investigated in KBr pellets by means of Fourier transform infrared (FT-IR) spectroscopy. In our focus of interest was to study the 3 and 10 μm (3000 and 1000 cm$^{-1}$, the O−H and silicate streching) bands of several CM2 chondrites. By investigating these signals the water content and the extent of hydration can be determined. In order to achieve this, development of a new pellet production method was essential. This technique facilitates the elimination of adsorbed water coming from the surrounding environment, which would complicate correct interpretation of the results.

**Experimental details.** Meteorite samples were examined in KBr pellets pressed in hydraulic press. The meteorite samples were previously mixed with KBr ($m_{sample}$ = 1.5 mg, $m_{KBr}$ = 300 mg) in a Retsch MM200 mixer mill for 15 minutes (mixing frequency $\nu_{mix}$ was set at 30 min$^{-1}$) then placed in a preheated pellet die ($T_{die,1}$ = 70 °C). It was then kept at 130 °C for 4 hours under vacuum ($p_{die}$ < 0.8 mbar). This was enough to get rid of the water adsorbed on the surface of the KBr but not to eliminate the water content of the silicates found in the meteorite grains. After this dehydration process pellets were pressed at this temperature and pressure for 60 minutes, then IR spectra of them were taken by a Bruker Vertex v70 infrared spectrometer.

**Theoretical background.** To assure that interpretation of the experimental results is correct, IR spectra calculated from optical constants ($n_i$, $k_i$) determined by the Laboratory Astrophysics Group in Jena have been compared to the experimental ones of olivine and meteorite samples. From the optical constants, the complex dielectric function ($\varepsilon_i$) and the extinction cross-section ($C_{ext,i}$) can be calculated. Three different models have been used: one assuming that particles are spheres, the other one take only the absorption (the imaginary part of the optical constant) into account and the CDE1 model developed by Fabian et al.[1] which assumes that the sample particles are ellipsoids with equal probability of all shapes. Models that consider not only the absorption but the scattering as well assume that the particle sizes are in the Rayleigh-limit ($2\pi r/\lambda \ll 1$ where $r$ is the particle radius)

**Evaluation of results.** Pellets of eight meteorite samples were made and their IR spectra were taken. These meteorites were (by increasing order of their water content): MIL 07700, EET 97522, EET 96029, LEW 85311, MCY 05230, LEW 90500, LEW 87022 and ALH 84029). Using this pellet production method almost all adsorbed water signals, which would otherwise overlap with the ones come from meteoritic water disappeared. The method, however, was not sufficient to get rid of the mesoporic and interlayer water molecules. Nonetheless, signal strengths of the phylloscilicate O−H band at 3 μm and the olivine:phyllosilicate band ratio at 10 μm were clearly showed the meteoritic water content therefore the extent of aqueous alteration. By comparing the calculated spectra with the experimental ones the best agreement can be observed when CDE1 model is used in the case of olivine. However, particle sizes in the meteorite samples were presumably too big therefore the agreement was worse.

**Conclusion, future goals.** Using this KBr pellet production method permits to investigate water content of aqueously altered chondritic meteorites. These experimental spectra have been compared to calculated ones and in the case of olivine they agreed well. Higher temperatures would allow to remove the mesoporic and interlayer water molecules as well, thus interpretation of spectra would become even easier. Size separation of the meteorite samples would allow the correct comparison of spectra calculated by CDE1 model with the experimental ones.

**Acknowledgment.** This work was funded by TÉT_12_FR-1-2013-0023 Grant.

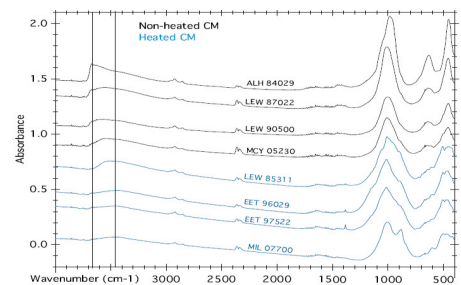

---